\documentclass[a4paper,12pt]{article}

\usepackage[dvips]{graphicx}
\usepackage{epsfig}

\begin{document}

\author{P. A. Hogan\thanks{E-mail : peter.hogan@ucd.ie} and S. O'Farrell\thanks{E-mail : shanefarrell ie@hotmail.com}\\
\small School of Physics,\\ \small University College Dublin,
Belfield, Dublin 4, Ireland}

\title{Generating Electromagnetic Waves from Gravity Waves in
Cosmology}
\date{PACS number(s): 04.30.Nk}
\maketitle

\begin{abstract}Examples of test electromagnetic waves on a
Friedmann--Lema\^itre--Robertson--Walker (FLRW) background are
constructed from explicit perturbations of the FLRW space--times
describing gravitational waves propagating in the isotropic
universes. A possible physical mechanism for the production of the
test electromagnetic waves is shown to be the coupling of the
gravitational waves with a test magnetic field, confirming the
observation of Marklund, Dunsby and Brodin [Phys. Rev. D{\bf 62},
101501(R) (2000)].
\end{abstract}
\thispagestyle{empty}
\newpage

\section{Introduction}\indent
Marklund, Dunsby and Brodin \cite{MDB} have made the important
observation that gravitational wave perturbations of
Friedmann--Lema\^itre--Robertson--Walker (FLRW) cosmological
models coupled to weak magnetic test fields can generate
electromagnetic waves. The magnitude of this effect has been
estimated in \cite{MDB}. They have demonstrated this phenomenon
using the gauge--invariant and covariant perturbation theory of
Ellis and Bruni \cite{EB}. This theory has also been used to
construct explicit solutions of the Ellis--Bruni perturbation
equations describing gravitational waves propagating in FLRW
universes \cite{HOS} (see also \cite{HE}) \emph{from a point of
view which differs significantly from that of} \cite{MDB}. The
histories of the wave fronts of the latter waves are particularly
simple families of null hypersurfaces which arise naturally in the
FLRW space--times. In this paper we demonstrate how test
electromagnetic waves can be constructed from these explicit
gravitational waves and how these electromagnetic waves can be
viewed as arising from the interaction of the gravitational waves
with a test magnetic field, thereby supporting the observation of
Marklund, Dunsby and Brodin.

We utilize a test magnetic field which, by its nature, does not
perturb the isotropic cosmology but is nevertheless a violation of
isotropy. The gravitational waves we use do perturb the isotropic
cosmology and this perturbation also breaks the isotropic symmetry
since the gravitational waves are uni--directional and thus at
each point of the isotropic cosmological model their histories
have a unique null propagation direction. The behavior of magnetic
fields in cosmological models has been extensively and carefully
studied from diverse physical viewpoints in \cite{T1}--\cite{BT}.

The gravitational waves which we utilize in this paper correspond
to the simplest (from a geometrical point of view) type of
gravitational radiation which can propagate in isotropic
cosmologies. The histories of their wave fronts are naturally
occurring null hypersurfaces in the isotropic cosmological models
and their propagation direction in these space--times is null,
geodesic and shear--free. It is thus of some interest to examine
the Marklund, Dunsby and Brodin observation in terms of them. As a
measure of the strength of the gravitational waves used in this
paper, and of the electromagnetic waves which result from their
interaction with a test magnetic field, we find that the
gravitational field of the gravitational waves (the perturbed Weyl
tensor) is proportional to $R^{-2}$ (see eq.(2.34)) where $R(t)$
is the scale factor of the isotropic universe, while the
electromagnetic field (see eq. (3.17)) is also proportional to
$R^{-2}$. The coefficients of $R^{-2}$ in both cases involve an
arbitrary analytic function whose appearance is a characteristic
of electromagnetic radiation (which is shear--free \cite{R} in the
optical sense that the null propagation direction in space--time,
given by the gradient of the function $\phi$ in (2.28) below, is
shear--free) and of shear--free (in the optical sense)
gravitational radiation \cite{RT}.

The interaction of gravitational waves and electromagnetic fields
has led to proposals for a mechanism to detect gravitational waves
\cite{L}--\cite{GS}. Our approach is influenced by the fundamental
paper by Szekeres \cite{S} on the interaction of gravitational
waves with matter and with electromagnetic waves. With the
inclusion of matter many modelling possibilities open up. For
example, a recent model for the generation of gravitational waves
from matter and electromagnetic waves can be found in \cite{BH}.

This paper is organized in the following way: in section 2 the
relevant Ellis--Bruni perturbation equations (for tensor
perturbations) are listed and the explicit solutions derived in
\cite{HOS} are summarized. Analogous test electromagnetic waves
having the same wave fronts as these gravitational waves will be
required and they are described in section 3. In section 4 the
construction of test electromagnetic waves of the type considered
in section 3 are derived from the gravitational waves of section
2. In addition in this section these electromagnetic waves are
shown to be capable of being interpreted as resulting from the
interaction of the gravitational waves with an explicit test
magnetic field. The paper ends with a brief discussion in section
5.

\setcounter{equation}{0}
\section{Gravity Waves in FLRW Universes}\indent
We make use of a four dimensional space--time manifold with a
metric tensor having components $g_{ab}$ in a local coordinate
system $\{x^a\}$. This manifold contains a preferred congruence of
time--like world--lines which are the integral curves of a vector
field having components $u^a$ in the coordinate system $\{x^a\}$
and which satisfies $u_a\,u^a=-1$. The electric and magnetic parts
of the Weyl tensor $C_{abcd}$ are defined respectively by
\begin{equation}\label{2.1}
E_{ab}=C_{apbq}\,u^p\,u^q\ ,\qquad H_{ab}=C^*_{apbq}\,u^p\,u^q\
.\end{equation}These are equivalent to $C_{abcd}$ and the star
superscript denotes the dual $C^*_{apbq}=\frac{1}{2}\eta
_{ap}{}^{rs}\,C_{rsbq}$ with $\eta _{abcd}=\sqrt{-g}\,\epsilon
_{abcd}$, $g={\rm det}(g_{ab})$ and $\epsilon _{abcd}$ the
Levi--Civita permutation symbol. We note that for the Weyl tensor
the left and right duals are equal. The energy--momentum--stress
tensor $T^{ab}=T^{ba}$ describing the matter distribution can be
decomposed with the respect to the vector field $u^a$ to read
\cite{E}
\begin{equation}\label{2.2}
T^{ab}=\mu\,u^a\,u^b+p\,h^{ab}+q^a\,u^b+q^b\,u^a+\pi ^{ab}\
,\end{equation}with $\mu$ the density of matter measured by the
observer with 4--velocity $u^a$, $h^{ab}=g^{ab}+u^a\,u^b$ the
projection tensor, $p$ the isotropic pressure, $q^a$ the energy
flow measured by the observer with 4--velocity $u^a$ and $\pi
^{ab}=\pi ^{ba}$ the anisotropic stress. Here
\begin{equation}\label{2.3}
q^a\,u_a=0\ ,\qquad \pi ^{ab}\,u_b=0\ ,\qquad \pi ^a{}_a=0\
.\end{equation}The covariant derivative $u_{a;b}$ is decomposed
into
\begin{equation}\label{2.4}
u_{a;b}=\omega _{ab}+\sigma _{ab}+\frac{1}{3}\theta\,h_{ab}-\dot
u_a\,u_b\ ,\end{equation}with the dot in general indicating
covariant differentiation in the direction of $u^a$ (and thus in
particular $\dot u^a=u^a{}_{;b}\,u^b$). Here
\begin{equation}\label{2.5}
\omega _{ab}=u_{[a;b]}+\dot u_{[a}\,u_{b]}\ ,\end{equation}is the
vorticity tensor (with $\omega _{ab}=-\omega _{ba}\ ,\omega
_{ab}\,u^b=0$ and square brackets denote skew symmetrization),
\begin{equation}\label{2.6}
\sigma _{ab}=u_{(a;b)}+\dot
u_{(a}\,u_{b)}-\frac{1}{3}\theta\,h_{ab}\ ,\end{equation}is the
shear tensor (with $\sigma _{ab}=\sigma _{ba}\ ,\sigma ^a{}_a=0\
,\sigma _{ab}\,u^b=0$ and round brackets denote symmetrization)
and
\begin{equation}\label{2.7}
\theta =u^a{}_{;a}\ ,\end{equation}is the expansion or contraction
scalar.

In the isotropic FLRW space--times with $u^a$ the 4--velocity of
matter we have $q^a=0$ and $\pi ^{ab}=0$ and thus (\ref{2.2})
specialises to the perfect fluid form
\begin{equation}\label{2.8}
T^{ab}=\mu\,u^a\,u^b+p\,h^{ab}\ ,\end{equation}with, in addition,
$h^b_a\,p_{,b}=0$ and $h^b_a\,\mu _{,b}=0$. We also have in this
case $\dot u^a=0$, $\omega _{ab}=0$, $\sigma _{ab}=0$ and
$h^b_a\,\theta _{,b}=0$ which has the effect of simplifying
(\ref{2.4}) to
\begin{equation}\label{2.9}
u_{a;b}=\frac{1}{3}\theta\,h_{ab}\ .\end{equation}In this case
$\theta$ satisfies the simplified Raychaudhuri equation
\begin{equation}\label{2.10}
\dot\theta+\frac{1}{3}\theta ^2=-\frac{1}{2}(\mu +3\,p)\
,\end{equation} and the equations $T^{ab}{}_{;b}=0$ reduce in this
case to the single equation
\begin{equation}\label{2.11}
\dot\mu +\theta\,(\mu +p)=0\ .\end{equation}The space--times are
now necessarily conformally flat and thus  $E_{ab}=0$ and
$H_{ab}=0$. The metric tensor $g_{ab}$ is given via the
line--element in the Robertson--Walker form
\begin{equation}\label{2.12}
ds^2=R^2(t)\,\frac{[(dx^1)^2+(dx^2)^2+(dx^3)^2]}{\left
(1+\frac{k}{4}\,r^2\right )^2}-dt^2\ ,\end{equation}with $R(t)$
the scale factor, $r^2=(x^1)^2+(x^2)^2+(x^3)^2$ and $k(=0,\ \pm
1)$ is the Gaussian curvature of the $t={\rm constant}$
space--like hypersurfaces. In these coordinates $u^a\partial
/\partial x^a=\partial /\partial t$ and $x^{\alpha}$ $(\alpha =1,
2, 3)$ are constant on each integral curve of the vector field
$\partial /\partial t$. In addition $\theta =\theta (t),\ \mu =\mu
(t)$ and $p=p(t)$ are obtained from the scale factor $R(t)$ as
$\theta =3\,\dot R/R$ with $\mu ,\ p$ given by (\ref{2.11}) along
with an equation of state while $R(t)$ is derived from the
Friedmann equation (whose time derivative in this case coincides
with (\ref{2.10}); see, for example \cite{E}).

In the Ellis--Bruni \cite{EB} theory perturbations of the FLRW
models in particular are described by the basic gauge invariant
variables $E_{ab},\ H_{ab},\ \sigma _{ab},\ \omega _{ab},\ \dot
u^a$, $h^b_a\,\theta _{,b},\ h^b_a\,\mu _{,b},\ h^b_a\,p_{,b},\
q^a$ and $\pi ^{ab}$, which are taken to be small of first order.
The differential equations determining these variables are
obtained by retaining first order small terms in the Ricci
identities, the Bianchi identities and the equations
$T^{ab}{}_{;b}=0$ with $T^{ab}$ given by (\ref{2.2}). There are
two exceptions to this procedure: the propagation equation for
$\theta$ along the integral curves of $u^a$ (Raychaudhuri's
equation), which follows from the Ricci identities, and the
propagation equation for $\mu$ along the integral curves of $u^a$,
which follows from $T^{ab}{}_{;b}=0$. These equations are
converted  into differential equations for the gauge invariant
variables  listed above by projecting their gradients orthogonal
to $u^a$ (see  \cite{EB}). For perturbations that exclusively
describe  gravitational waves propagating through FLRW universes
we only  require the variables $E_{ab},\ H_{ab},\ \sigma _{ab},\
\pi ^{ab}$ (the so--called ``tensor" perturbations), with the
remaining gauge invariant variables vanishing. The linear
equations determining these first order quantities are \cite{HOS}
\begin{equation}\label{2.13}
\dot\sigma _{ab}+\frac{2}{3}\,\theta\,\sigma
_{ab}-\frac{1}{2}\,\pi _{ab}+E_{ab}=0\ ,\end{equation}and
\begin{equation}\label{2.14}
H_{ab}=-\sigma _{(a}{}^{g;c}\,\eta _{b)fgc}\,u^f\
,\end{equation}from the Ricci identities and
\begin{eqnarray}\label{2.15}
E^{bd}{}_{;d}&=&-\frac{1}{2}\,\pi ^{bd}{}_{;d}\ ,\\
H^{bd}{}_{;d}&=&0\ ,\\
\dot E^{bt}+\theta\,E^{bt}&=&-u_r\,H^{(b}_{s;d}\,\eta
^{t)rsd}-\frac{1}{2}\,\dot\pi ^{bt}-\frac{1}{6}\,\theta\,\pi
^{bt}\ ,\\
\dot H^{bt}+\theta\,H^{bt}&=&u_r\,E^{(b}_{s;d}\,\eta
^{t)rsd}-\frac{1}{2}\,\eta ^{(b}{}_{rad}\,\pi ^{t)a;d}\,u_r\
,\end{eqnarray}from the Bianchi identities. To obtain simple
solutions of these equations we look for solutions which have an
arbitrary dependence on a scalar function $\phi (x^a)$ by writing
\begin{equation}\label{2.16}
\sigma _{ab}=s_{ab}\,F(\phi )\ ,\qquad \pi _{ab}=\Pi _{ab}\,F(\phi
)\ .\end{equation}We will then substitute these into (\ref{2.13})
and (\ref{2.14}) to obtain expressions for $E_{ab}$ and $H_{ab}$
which depend on $F$ and its derivative with respect to $\phi$.
These, along with (\ref{2.16}), are then substituted into
(2.15)--(2.18) to arrive at differential equations for $s_{ab}$
and $\Pi _{ab}$. We note that
\begin{equation}\label{2.17}
s_{ab}=s_{ba},\ s_{ab}\,u^b=0,\ g^{ab}\,s_{ab}=s^a{}_a=0\ ,\end
{equation}with similar equations holding for $\Pi _{ab}$. When the
substitutions indicated above are carried out the consistency of
the resulting equations requires \cite{HOS}
\begin{equation}\label{2.18}
g^{ab}\,\phi _{,a}\,\phi _{,b}=0\ ,\end{equation}(the comma
denoting partial differentiation with respect to $x^a$) so that
the hypersurfaces $\phi (x^a)={\rm constant}$ are null, and the
following consistent equations (also consistent with (\ref{2.17}))
must hold:
\begin{equation}\label{2.19}
s^{ab}\,\phi _{,b}=0\ ,\qquad \Pi ^{ab}\,\phi _{,b}=0\
,\end{equation}along with
\begin{equation}\label{2.20}
s^{ab}{}_{;b}=0\ ,\qquad \Pi ^{ab}{}_{;b}=0\ ,\end{equation}the
propagation equation for $s^{ab}$ along the generators of the null
hypersurfaces $\phi (x^a)={\rm constant}$,
\begin{equation}\label{2.21}
s^{ab;c}\,\phi _{,c}+\left (\frac{1}{2}\phi
^{,d}{}_{;d}-\frac{1}{3}\theta\,\dot\phi\right
)\,s^{ab}=-\frac{1}{2}\dot\phi\,\Pi ^{ab}\ ,\end{equation}and the
wave equation,
\begin{equation}\label{2.22}
s^{ab;d}{}_{;d}-\frac{2}{3}\theta\,\dot s^{ab}+\left
(-\frac{1}{3}\theta ^2+\frac{3}{2}p-\frac{1}{6}\mu\right
)s^{ab}=-\dot\Pi ^{ab}-\frac{2}{3}\theta\,\Pi ^{ab}\
.\end{equation}The null hypersurfaces $\phi (x^a)={\rm constant}$
are the histories in the FLRW space--times of the wave fronts of
the gravitational waves described by these perturbations (see
\cite{HOS}, \cite{HE}). To exhibit explicit solutions we first
choose these null hypersurfaces. To obtain some naturally occuring
null hypersurfaces which will lead to surveyable solutions we
start by writing the Robertson--Walker line--element (\ref{2.12})
in the form \cite{H}
\begin{equation}\label{2.23}
ds^2=R^2(t)\{dx^2+p_0^{-2}f^2(dy^2+dz^2)\}-dt^2\ ,\end{equation}
with $p_0=1+(K/4)(y^2+z^2),\ K={\rm constant},\ f=f(x)$. The
following cases are allowed: (i) if $k=+1$ then $K=+1$ and
$f(x)=\sin x$; (ii) if $k=0$ then $K=0, +1$ with $f(x)=1$ when
$K=0$ and $f(x)=x$ when $K=+1$; (iii) if $k=-1$ then $K=0,\ \pm 1$
with $f(x)=\frac{1}{2}\,e^x$ when $K=0$, $f(x)=\sinh x$ when
$K=+1$ and $f(x)=\cosh x$ when $K=-1$. The details of these
special cases are given in \cite{HOS}. The following equations are
satisfied in the cases (i)--(iii):
\begin{equation}\label{2.24'}
f''+k\,f=0\ ,\qquad (f')^2+k\,f^2=K\ ,\end{equation}with the prime
denoting differentiation with respect to $x$. A convenient family
of null hypersurfaces is given by
\begin{equation}\label{2.24}
\phi (x^a)=x-T(t)={\rm constant}\ ,\end{equation}with
$dT/dt=R^{-1}$.

Using the null hypersurfaces (\ref{2.24}) and the equations
(2.23)--(2.25) for $s^{ab},\ \Pi ^{ab}$ subject to (2.20) (and
similar conditions on $\Pi ^{ab}$) and (2.22) yields solutions in
the form
\begin{eqnarray}\label{2.25}
s^{ab}&=&\bar s\,m^a\,m^b+s\,\bar m^a\,\bar m^b\ ,\\
\Pi ^{ab}&=&\bar\Pi\,m^a\,m^b+\Pi\,\bar m^a\,\bar m^b\
,\end{eqnarray}with $m^a$ given by the 1--form
$m_a\,dx^a=R\,p_0^{-1}f\,d\zeta /\sqrt{2}$ with $\zeta =y+iz$ (and
thus $m_a\,m^a=0=\bar m_a\,\bar m^a$ and $m_a\,\bar m^a=+1$) and
the bar denoting complex conjugation. In addition
\begin{eqnarray}\label{2.26}
\bar s&=&-\frac{p_0^2}{R\,f}\,{\cal G}(\zeta , x, t)\ ,\\
\bar\Pi &=&-\frac{2\,p_0^2}{R^2f}\,(D{\cal G}+\dot R\,{\cal G})\
.\end{eqnarray}Here ${\cal G}$ is a complex analytic function of
its argument and $D=\partial /\partial x+R\,\partial /\partial t$.
Also ${\cal G}$ satisfies
\begin{equation}\label{2.27}
D^2{\cal G}+k\,{\cal G}=0\ ,\end{equation}with $k=0, \pm 1$ and
this can easily be solved with each case involving two arbitrary
complex analytic functions,  of $\zeta$ and $x-T=\phi$, of
integration. The corresponding perturbed Weyl tensor is given by
\begin{equation}\label{2.28}
E^{ab}+iH^{ab}=-\frac{2\,p_0^2}{R^2f}\,\frac{\partial}{\partial
x}({\cal G}\,F(\phi ))\,m^a\,m^b\ ,\end{equation}which is a Petrov
type N Weyl tensor with degenerate principal null direction $\phi
_{,a}$, confirming the interpretation of the perturbations of the
FLRW space--times described here as being due to gravitational
waves having the null hypersurfaces $\phi ={\rm constant}$ as the
histories of their wave fronts. Further properties of these waves
are discussed in \cite{HOS}.

\setcounter{equation}{0}
\section{Electromagnetic Waves in FLRW Universes}\indent
Electromagnetic test fields on the space--time described at the
beginning of section 2 are encoded in a skew--symmetric tensor
with components $F_{ab}=-F_{ba}$ having electric and magnetic
parts defined by \cite{E} \begin{equation}\label{3.1}
E_a=F_{ab}\,u^b\ ,\qquad H_a=F^*_{ab}\,u^b\ ,\end{equation}with
$F^*_{ab}=\frac{1}{2}\eta _{ab}{}^{rs}\,F_{rs}$. The vectors $E_a,
H_a$ are equivalent to a knowledge of $F_{ab}$ with
\begin{equation}\label{3.2}
F_{ab}=u_a\,E_b-u_b\,E_a-\eta _{abcd}\,u^c\,H^d\ .\end{equation}If
these are source--free electromagnetic test fields on the FLRW
space--times then they must satisfy Maxwell's equations in the
form \cite{E}
\begin{equation}\label{3.3}
E^a{}_{;a}=0\ ,\qquad H^a{}_{;a}=0\ ,\end{equation}and
\begin{eqnarray}\label{3.4}
\dot E^a+\frac{2}{3}\theta\,E^a&=&-\eta ^{abed}\,u_b\,H_{e;d}\ ,\\
\dot H^a+\frac{2}{3}\theta\,H^a&=&\eta ^{abed}\,u_b\,E_{e;d}\
.\end{eqnarray}To obtain solutions analogous to the gravitational
waves described in section 2 we first introduce a 4--potential
$\sigma ^a$ with \cite{HE2}
\begin{equation}\label{3.5}
\sigma ^a\,u_a=0\ ,\qquad \sigma ^a{}_{;a}=0\ ,\end{equation}and
\begin{equation}\label{3.6}
F_{ab}=\sigma _{b;a}-\sigma _{a;b}\ .\end{equation}Using the Ricci
identities along with the conformal flatness of the FLRW
space--times we have
\begin{equation}\label{3.7}
\sigma _{a;dc}-\sigma _{a;cd}=\frac{2}{3}\mu\,g_{a[c}\,\sigma
_{d]}+(\mu +p)\,u_a\,u_{[c}\,\sigma _{d]}\ .\end{equation}This
helps to establish that (\ref{3.3}) are now satisfied
automatically. In addition (\ref{3.4}) reduces to the wave
equation
\begin{equation}\label{3.8}
\sigma ^{a;d}{}_{;d}=\frac{1}{2}(\mu -p)\,\sigma ^a\
,\end{equation}and (3.5) is automatically satisfied. The equations
to be satisfied by $\sigma ^a$ are (\ref{3.5}) and (\ref{3.8}).
Following (\ref{2.16}) we look for solutions of the form
\begin{equation}\label{3.9}
\sigma ^a=s^a\,F(\phi )\ ,\end{equation} with $F$ an arbitrary
function of $\phi (x^a)$. With $s^a\neq 0$ we arrive again at
(\ref{2.18}) along with
\begin{equation}\label{3.10}
s^a{}_{;a}=0\ ,\qquad s^a\,\phi _{,a}=0\ .\end{equation}We also
obtain the propagation equation for $s^a$ along the generators of
the null hypersurfaces $\phi (x^a)={\rm constant}$,
\begin{equation}\label{3.11}
s^{a;b}\,\phi _{,b}+\frac{1}{2}\phi _{,d}{}^{;d}\,s^a=0\
,\end{equation} and the wave equation,
\begin{equation}\label{3.12}
s^{a;d}{}_{;d}=\frac{1}{2}(\mu -p)\,s^a\ .\end{equation}Solving
these with $\phi (x^a)$ given by (\ref{2.24}) we find
\begin{equation}\label{3.13}
s^a=\bar s_0\,m^a+s_0\,\bar m^a\ ,\end{equation}with $m^a$ given
following (\ref{2.25}) and
\begin{equation}\label{3.14}
\bar s_0=\frac{p_0}{R\,f}\,{\cal F}(\zeta , x, t)\
,\end{equation}with ${\cal F}$ a complex analytic function
required to satisfy
\begin{equation}\label{3.15}
D{\cal F}=0\ ,\end{equation}with the operator $D$ defined
following (2.32). Thus ${\cal F}={\cal F}(\zeta ,\ x-T)$ and is
otherwise arbitrary. The corresponding electric and magnetic
fields (\ref{3.1}) are given via
\begin{equation}\label{3.16}
E^a+iH^a=\frac{2\,p_0}{R^2f}\,\frac{\partial}{\partial x}({\cal
F}\,F)\,m^a\ ,\end{equation}analogous to (\ref{2.28}). The
radiative nature of this electromagnetic field, with propagation
direction $\phi _{,a}$, is evident from (\ref{3.16}).

\setcounter{equation}{0}
\section{Electromagnetic Waves from Gravity Waves}\indent
We now demonstrate how electromagnetic waves of the type described
in section 3 can be constructed from the gravitational waves given
in section 2. The construction is not unique (see the discussion
in the next section) but it can be given the interpretation of
arising from the coupling of the gravitational waves of section 2
with a test magnetic field in agreement with \cite{MDB}.

The gravitational waves of section 2 are obtained from a complex
analytic function ${\cal G}(\zeta , x, t)$ which satisfies
(\ref{2.27}). We see from the argument in section 3 that the
electromagnetic waves given there are obtained from a complex
analytic function ${\cal F}(\zeta , x, t)$ which satisfies
(\ref{3.15}). Given ${\cal G}$ satisfying (\ref{2.27}) and $f(x)$
satisfying (\ref{2.24'}) we define
\begin{equation}\label{4.1}
{\cal F}(\zeta , x, t)=f\,D{\cal G}-f'\,{\cal G}\ .\end{equation}
Clearly this function satisfies (\ref{3.15}) and so ${\cal
F}={\cal F}(\zeta , x-T)$. With this choice of ${\cal F}$
electromagnetic waves in section 3 corresponding to the
gravitational waves of section 2 are described by the 4--potential
(\ref{3.9}) with $s^a$ given by (\ref{3.13}) and (\ref{3.14}).

A source--free test magnetic field on the FLRW space--times must
satisfy (3.3)--(3.5) with $H_a\neq 0$ and $E_a=0$. A simple
example of such a field is given by
\begin{equation}\label{4.5}
H_a=\frac{\lambda _a}{R\,f^2}\ ,\qquad \lambda _a=h^b_a\,\phi
_{,b}\ .\end{equation}This field can be derived from a
4--potential in the manner of (\ref{3.6}) with (\ref{3.5}). In
this case the potential is given by the 1--form
\begin{equation}\label{4.6}
{}_H\sigma _a\,dx^a=\frac{y\,dz-z\,dy}{2\,p_0}\ ,\end{equation}or
equivalently by the covariant vector
\begin{equation}\label{4.7}
{}_H\sigma ^a=\frac{i(\zeta\,\bar
m^a-\bar\zeta\,m^a)}{2\sqrt{2}\,R\,f}\ ,\end{equation}with $m^a$
given following (2.30). We note in passing that any source--free
magnetic test field on the FLRW space--times is given by
$H_a=R^{-1}q_{,a}$ with $q_{,a}u^a=0$ and $g^{ab}q_{,a;b}=0$ and
(\ref{4.5}) corresponds to $q=q(x)$ given by $dq/dx=f^{-2}$.

Observing that the function ${\cal F}$ in (\ref{4.1}) is linear in
the arbitrary analytic function ${\cal G}$ and its derivative
$D{\cal G}$ we first note from (2.29)--(2.32) that
\begin{equation}\label{4.8}
s^{ab}=-\frac{p_0^2}{R\,f}\,{\cal
G}\,m^a\,m^b-\frac{p_0^2}{R\,f}\,\bar{\cal G}\,\bar m^a\,\bar m^b\
,\end{equation}and
\begin{equation}\label{4.9}
\Pi
^{ab}-\frac{2}{3}\,\theta\,s^{ab}=-\frac{2\,p_0^2}{R^2f}\,D{\cal
G}\,m^a\,m^b-\frac{2\,p_0^2}{R^2f}\,D\bar{\cal G}\,\bar m^a\,\bar
m^b\ .\end{equation}Hence using (\ref{4.7}) we arrive at
\begin{equation}\label{4.10}
-2\,p_0^{-1}R\,f\,f'\,s^{ab}{}_H\sigma _b-p_0^{-1}R^2f^2\left (\Pi
^{ab}-\frac{2}{3}\theta\,s^{ab}\right ){}_H\sigma
_b=\frac{p_0}{R\,f}\,({\cal J}\,m^a+\bar{\cal J}\,\bar m^a)\
,\end{equation}with
\begin{equation}\label{4.11}
{\cal J}=\frac{i\zeta}{\sqrt{2}}\,{\cal F}\ ,\end{equation}and
${\cal F}$ is given by (\ref{4.1}). The right hand side of
(\ref{4.10}) is the 4--potential of a radiative Maxwell test field
on the FLRW space--times of the type described in section 3. The
left hand side of (\ref{4.10}) provides a Marklund--Dunsby--Brodin
physical interpretation of the origin of this electromagnetic
field as a coupling of the gravitational waves described by
$s^{ab}$ and $\Pi ^{ab}$ with the magnetic test field described by
the 4--potential ${}_H\sigma ^a$, due to the appearance of the
products $s^{ab}{}_H\sigma _b$ and $\Pi ^{ab}{}_H\sigma _b$.

\setcounter{equation}{0}
\section{Discussion}\indent
We make two observations regarding the electromagnetic waves
constructed in section 4 from the gravitational waves in section
2. The first observation concerns the lack of uniqueness of the
resulting electromagnetic waves. To see this we note that if
${\cal G}$ and ${\cal H}$ are two complex analytic functions
satisfying (\ref{2.27}) then
\begin{equation}\label{5.1}
{\cal F}(\zeta , x, t)=k\,{\cal H}\,{\cal G}+D{\cal H}\,D{\cal G}\
,\end{equation}satisfies (\ref{3.15}). In view of (\ref{2.24'}) a
choice of ${\cal H}$ is simply ${\cal H}=f(x)$. In this case
\begin{equation}\label{5.2}
{\cal F}(\zeta , x, t)=k\,f\,{\cal G}+f'D{\cal G}\
.\end{equation}It is easy to see from (\ref{3.16}) that these
waves are distinct from those obtained using ${\cal F}$ given in
(\ref{4.1}). In addition we note that there are no waves of this
type for the case (ii) with $k=K=0$ given following (\ref{2.23}),
in contradistinction to the example (\ref{4.1}). On the other hand
since (\ref{5.1}) is a linear combination of ${\cal G}$ and
$D{\cal G}$ these electromagnetic waves also lend themselves to
the interpretation of arising from the coupling of the
gravitational waves of section 2 to the test magnetic field given
in section 4.

The significance of utilizing a test magnetic field rather than a
test electric field in (\ref{4.5}) is clarified by the following
observation: A source--free test electric field given by (3.1),
(3.6) and (3.7) with $H_a=0$ has a potential 1--form of the form
${}_E\sigma _a\,dx^a=a_1\,dx+a_2\,dy+a_3\,dz$ with $a_{\alpha}\
(\alpha =1, 2, 3)$ functions of $x, y, z, t$. Maxwell's equations
(3.3)--(3.5) with $H_a=0$ imply that ${}_E\sigma ^a=R^{-1}l(x, y,
z)\,u^a$, for some function $l$, up to a gauge transformation. The
scalar products of this 4--potential with $s^{ab}$ and $\Pi ^{ab}$
are therefore zero.

Finally we remark that astrophysical gravitational waves are
generally low frequency and in universes containing ionized matter
the corresponding electromagnetic waves propagate poorly. In our
treatment no mention has been made of the frequency of the
resulting electromagnetic waves with 4--potential (4.7). This
would be introduced via a Fourier analysis of the arbitrary
analytic function ${\cal G}$. To further study this
electromagnetic radiation and follow the dissipative consequences
of its interaction with the matter and gravitational radiation
would require calculating its perturbative effect on the
cosmological model. In this paper this electromagnetic radiation
is weak but is still considered a test field.

\setcounter{equation}{0}
\section{Acknowledgement}\indent
SOF wishes to thank IRCSET for financial support.

\end{document}